# A stem-cell ageing hypothesis on the origin of Parkinson's disease

André X. C. N. Valente[1,2,*], Jorge A. B. Sousa[2], Tiago Fleming Outeiro[3,4], Lino Ferreira[1,5]

[1]Center for Neurosciences and Cell Biology, University of Coimbra, Coimbra 3004-517, Portugal

[2]Systems Biology Group, Biocant, Cantanhede 3060-197, Portugal

[3]Cell and Molecular Neuroscience Unit, Instituto de Medicina Molecular, Lisboa, Portugal

[4]Instituto de Fisiologia, Faculdade de Medicina, Universidade de Lisboa, Av. Prof. Egas Moniz, 1649-028 Lisboa, Portugal

[5]Biomaterials and Stem Cell Research Group, Biocant, Cantanhede 3060-197, Portugal

[*]**Corresponding author:**

André X. C. N. Valente

Center for Neurosciences and Cell Biology, University of Coimbra, Coimbra 3004-517, Portugal

Email: andre.valente@biocant.pt
Phone: +351 231 419 040
Fax: +351 231 419 049

## Abstract

A transcriptome-wide blood expression dataset of Parkinson's disease (PD) patients and controls was analyzed under the hypothesis-rich mathematical framework. The analysis pointed towards differential expression in blood cells in many of the processes known or predicted to be disrupted in PD. We suggest that circulating blood cells in PD patients can be in a full-blown PD-expression state. We put forward the hypothesis that sporadic PD can originate as a case of hematopoietic stem cell/differentiation process expression program defect and suggest this research direction deserves further investigation.

**Introduction**

Parkinson's disease (PD) is the second most common neurodegenerative disorder, after Alzheimer's disease (1). With the increasing life-expectancy, the number of worldwide affected individuals is expected to double from ~5 million in 2005 to ~8 million by 2030 (2). The majority of PD cases are sporadic, with only 5-10% of cases presumed to have a well-defined singular genetic cause. In spite of significant insights and hypotheses proposed in recent years, the origin (etiology) of sporadic PD remains undetermined (1).

A landmark study has reported genome-wide expression data on RNA extracted from whole blood of 50 predominantly early-stage PD patients (mean Hoehn and Yahr stage 2.3, range 1-4) and 55 age-matched controls (3). The study data analysis focused on producing a candidate blood-based early diagnostic composite biomarker for PD, but not on any potential implications of the data towards a renewed biological understanding of PD and its etiology. In fact, many of the genes it identified as blood PD biomarkers have no known role in PD pathogenesis (3).

Recently, we developed a novel mathematical framework, named hypothesis-rich framework, for analyzing problems that are complex in the sense of involving a large number of a priori potentially relevant hypothesis, facts and data (4). The framework is ideal for addressing systems biology problems where there is potential to jointly leverage a diversity of existing biological knowledge and quantitative data on a large number of variables. Having significant prior knowledge, which has accumulated through the vast work on Parkinson's disease (PD) until the present day, and having a large number of biological variables in the 22 283 gene expression probe readings, the hypothesis-rich framework is well suited for profiling and interpreting the PD blood gene expression data.

Herein, we analyzed the PD blood gene expression dataset previously reported, now under the hypothesis-rich mathematical framework. We constructed a composite biomarker, based on the gene expression data, for distinguishing PD individuals from non-PD individuals. The composite biomarker showed an across-the-board presence of genes known to play a role in PD and neurodegenerative processes. We propose this raises the prospect of an integral presence of PD in the blood. We hypothesize that a hematopoietic stem-cell defect can be at the origin of sporadic PD.

**Results**

In settings where there are large number of candidate biomarker laws, noisy data and a limited number of observed cases, a fundamental obstacle arises in that many candidate laws will diagnose correctly the observed cases just by chance. This is the so called over-fitting, or under-determination issue. Our PD classifier problem falls in this regime: i) we are searching for a law potentially based on 22 283 probe readings; ii) we have data on just 105 patients and iii) noise can be a significant issue in microarray experiments (5). The hypothesis-rich approach to cope with this fundamental obstacle consists in leveraging biological knowledge and insights to group candidate laws into Rationality Classes (RCs). Separating candidate laws into RCs enhances the statistical ability to circumvent over-fitting, increasing the chance of finding good laws, by way of finding favorable RCs. Technically, this stems from different RCs having different True Distance Distribution and Correlation Structures, due to their distinct etiology and rationale (4). We refer the reader to another article (4) for the underlying theory and report here in *Methods* only how the technique was implemented in this PD biomarker problem.

The blood gene expression-based final composite PD biomarker law is shown in Table 1. The biomarker demonstrated a positive, predictive ability with a Receiver Operating Characteristic (ROC) area under the curve (AUC) of 0.59±0.07 (see *Methods*). Interestingly, most of the 14 laws that compose the PD biomarker turned out to be centered around genes previously associated with PD and/or other neurodegenerative pathologies:

1. CLTB: This gene, encoding clathrin, is associated with vesicle-mediated transport, especially in endocytosis. CLTB has been implicated in dopamine transporter endocytosis (6) and shown in a recent PD mouse model to be down-regulated in the striatum region of the brain (7). Both our biomarker and that of Scherzer et al. (3) associate high expression of CLTB in the blood with the presence of PD.
2. FPR3: This gene belongs to the formyl peptide receptor family (FPR1, FPR2, and FPR3) of G-protein coupled receptors. FPRs have been implicated in the pathogenesis of Alzheimer's disease (AD) and prion diseases (8). In addition, FPRs can activate microglia (9) which has been observed in brains from patients with PD (10). Both our analysis and that of Scherzer et al. (3) found FPR3 to be over-expressed in the blood of PD patients.

3. ITGA2*ITGB1*CD47 [Integrin Complex (11)]: Integrins have been shown to mediate the interaction between microglia, the resident macrophages of the Central Nervous System, and the fibrillar beta-amyloid plaques found in the brains of Alzheimer's disease patients (12). In our analysis, high values of the gene expression product of the integrin complex biomarked the Alzheimer's and other neurodegenerative diseases (A) group (vs. the Parkinson's disease (P) group and the healthy (H) group).
4. UBOX5*PRPF19: PRPF19 and UBOX5 have been described to have ubiquitin ligase activity (13). A high-throughput data analysis by Hourani et al. (14) included PRPF19 in a set of 35 genes differentially expressed in the left brain hemisphere of mice with induced PD compared to normal controls. We are not aware of studies specifically associating UBOX5 and neurodegenerative processes, beyond its generic ubiquitin ligase activity (15). Our analysis shows this gene expression product as being under-expressed in the A group (vs. the P and H groups).
5. CLTB*KIF5B: KIF5B, a member of the kinesin family, is associated with the transport of peptide-containing vesicles to neuron terminals (16). In a genome-wide expression profiling of substantia nigra dopamine neurons in PD patients vs. controls, KIF5B was identified as one of the PD differentially under-expressed genes (16). CLTB has been discussed above. Our biomarker associates high values of this gene expression product with the presence of PD.
6. PTGS2*CYBB: PTGS2 encodes the inducible form of cyclooxygenase (COX-2). Teismann et al. demonstrated that PTGS2 inhibition averts the formation of the oxidant species dopamine-quinone, which has been implicated in the pathogenesis of PD (17). A genetic association study showed that polymorphisms of the PTGS2 gene predispose to AD (18). In mice over-expressing the Swedish mutation of the amyloid precursor protein as a model of AD, it was found that lack of CYBB averted the development of oxidative stress, cerebro-vascular dysfunction and behavioral deficits (19). This gene expression product showed lower values in the H group vis-à-vis in the P and A groups.
7. GPX4: This gene encodes glutathione peroxidase 4 and is involved in inflammation and in the oxidative stress response (20). GPX4 is a target gene of DJ-1 (21), a protein associated with familial cases of PD (22). Blackinton et al., although not detecting changes in mRNA expression levels of GPX4, observed an increase in its protein expression levels in the cortex of PD patients vs. controls (21). GPX4 is also a key antioxidant against lipid peroxidation, which was shown to be an early event in AD (23). Our analysis showed GPX4 being expressed at lower levels in the A group.
8. PTGS2: Already discussed above. Our analysis showed PTGS2 being expressed at higher levels in the A group.
9. UCHL1: Mutations in UCHL1 are associated with familial cases of PD (24). UCHL1 hydrolyzes a peptide bond at the C-terminal glycine of ubiquitin. It was found to be decreased in the H group in our analysis.
10. ST13: The protein encoded by this gene is a cofactor of chaperone heat-shock protein HSP70 (25). Depletion of HSP70 has been associated with the formation of Lewy bodies in PD patients (26). In vitro and in vivo experiments, the latter using Caenorhabditis elegans, indicate that a lower expression level of ST13 could facilitate depletion of HSP70 (27). Both our biomarker and that of Scherzer et al. (3) associate high expression levels of ST13 in the blood with the H group.
11. ACTB: This gene, encodes beta-actin. Actins are highly conserved proteins involved in cell structure and motility (28). Beta-actin, in particular, was found to regulate platelet nitric-oxide synthase 3 activity (29). There is evidence for the presence of Beta-actin mRNAs within developing dendritic and axonal growth cones (30). A high-throughput study compared gene expression in cerebral cortices of AD patients with that in cerebral cortices of non-demented controls containing abundant amyloid-plaques (31). It found ACTB to be the second most differentially over-expressed gene in the AD patients. However, at the time, follow-up RT-PCR analysis did not corroborate this differential expression. Our biomarker associates high expression levels of ACTB with the A group.
12. SMAD3*STRAP: SMAD3 and STRAP are involved in transforming growth factor β signaling. (TGF-β) (32). TGF-β levels have been reported as elevated in the striatum and in ventricular cerebrospinal fluid in PD (33; 34). In the blood, this gene expression product showed higher values in the H group.
13. UBOX5*CUL4B: UBOX5 and CUL4B have been described to have ubiquitin ligase activity (13). We are not aware of studies specifically associating UBOX5 or CUL4B and PD, beyond the general fact that the ubiquitin proteasome system has been widely implicated in protein accumulation in neurodegeneration (15). In the blood, this gene expression product showed higher values in the P group.

14. SMAD3: Already discussed above. Our analysis showed SMAD3 being expressed at higher levels in the H group.

**Discussion**

From the standpoint of serving as an early PD diagnostic tool, our composite biomarker demonstrated a weaker predictive ability than that of the biomarker proposed and evaluated by Scherzer et al. (AUC of 0.69) (3). Although at present noise remains a significant factor in genome-wide expression measurements (5), there is intensive ongoing research to address this limitation, centered both on microarray technology and on emerging techniques such as deep sequencing (35). Whether reductions in such technical noise will reveal any of these PD blood biomarkers to be powerful enough for the practical diagnosis of PD, or whether their limited predictive power is intrinsically biological, is an open question to be answered in the coming years. However, most interestingly, our blood gene expression biomarker already shows a significant, across-the-board presence of genes and processes known to play a role in PD and in neurodegenerative processes. Differential expression in blood cells in many of the processes known or predicted to characterize PD in neurological tissues, opens the possibility that, more than just exhibiting side-effects of the presence of PD in neuronal tissues, circulating blood cells in PD patients may be, at least expression-wise, in a full state of PD, much in the same way as affected neurons are. In fact, several studies, when taken as a whole, give further credibility to this possibility. Cytoplasmic hybrid ("cybrid") models of PD, in which donor mtDNAs from PD patients' platelets are introduced into and expressed in neural tumor cells with identical nuclear genetic and environmental backgrounds, demonstrate many abnormalities in which increased oxidative stress drives downstream antioxidant response and cell death activating signaling pathways (36). PD cybrids spontaneously form Lewy bodies and Lewy neurites, linking platelet mtDNA expression to neuropathology (37). Furthermore, they show both reduced mitochondrial respiration and impaired organelle transport in processes (36; 38). Thus, PD cybrids demonstrate that mitochondria from at least one blood cell type, contain the gene expression programs which are necessary for the expression of several central features of the disease. In addition, peripheral blood lymphocytes (PBL) from PD patients display altered densities of D1 and D2 dopamine receptors (39), further supporting our hypothesis. We believe the possibility of an integral presence, at least expression-wise, of PD in circulating blood cells should be considered. It could have deep implications on the open question of the etiology of sporadic PD (the word sporadic will be subsumed henceforth) and we discuss these next.

One commonly held generic view is that an environmental agent (or agents), possibly exacerbated by genetic and/or age-related vulnerabilities, is at the root of PD (40). Proposed candidate agents include environmental poisons, such as pesticides and metals (40), and yet unidentified neurotropic pathogens, such as viruses (41) or a prion-like protein (42). Usually suggested points of induction, or routes of entry for such a pathogen, are the peripheral olfactory system (42) and the gastrointestinal tract (43), due to the known association of these systems with early clinical symptoms and also Lewy body pathology in the latter case. These two candidate routes have recently been combined into a dual-hit theory (41). The theory proposes the pathogen, possibly a virus, would initially enter the body via the nasal route, and then via being swallowed in saliva and mucus, cross the stomach wall. Oxidative stress effects spur another major research direction on the origin of PD (44). Finally, regardless of the ultimate trigger, the criticality of the neuro-inflammatory response in the PD onset process is also a matter of active debate (45).

Taking our observation as a starting point, we put forward an alternative hypothesis on the nature and etiology of PD (Fig.1). We first suggest that, at its root, PD is best defined as a characteristic deviation from normality in the expression program of cells. We call it the PD-expression state, or PD-state for short. Thus, we propose that circulating blood cells in PD patients are in a full PD-state. The PD-state would therefore be a generic cell state, not specific to neuronal cells. However, as also argued by others (46), due to the particular critical role that some expression programs play in neuronal tissues, the PD-state is catastrophic in them, leading to the observed neuronal-associated pathology. A crucial question then becomes whether the PD-state is propagated from neuronal cells to blood cells, or vice-versa. That is, where does it originate? Recent studies show PD pathological signs, such as Lewy bodies and Lewy neurites, being propagated to healthy neuronal grafts in PD patients only over a time-scale on the order of a decade (47; 48). Considering, by comparison, the much shorter life-time of most blood cells (the lifespan of red blood cells and platelets is 127 (49) and 4.4 (50) days, respectively), we argue that it is comparatively more realistic that the PD-state originated in the blood cells.

PD is markedly age associated, with only 4% of PD cases diagnosed in the United States occurring before age 50 (2). There is evidence that hematopoietic stem cells (HSCs) age, showing an altered cell surface phenotype and changes in metabolic activity and gene expression (51; 52). Recent studies demonstrated that this ageing process is a consequence of accumulation of DNA damage (53). These lesions can be propagated to daughter stem cells and to downstream lineages through the processes of self-renewal and differentiation. We propose that the PD-state acquired by blood cells could be a case of hematopoietic stem cell ageing. Under this premise, circulating endothelial progenitor cells, which undergo endothelial cell differentiation under appropriate inductive signals and form neovessels (54), become a candidate vehicle for propagation of the PD-state to other cells in the human body.

An early sign of PD is impaired sense of smell (42). The olfactory bulb was recently reported as being a site of continuous stem cell based tissue regeneration (55). Gastrointestinal dysfunction is another early sign of PD reported as much as 10 years before motor symptoms appear (56). Gastrointestinal function is very sensitive to the proper function of intestinal epithelial cells (57), which are replenished by local adult stem cells with tissue turnover in under 7 days (58). Assuming the ability of HSCs to propagate the PD-state to adult stem cells via circulating progenitor cells, would explain the early PD symptoms in sites of very active stem cell based tissue regeneration. The initial propagation of the PD-state to these tissues could be due to rapid self-renewing and stem cell plasticity facilitating stem cell reprogramming by the endothelial cells derived from the circulating progenitor cells (59; 60). Of note, α-synuclein has been shown to be expressed in endothelial cells (61). Alternatively, the niche stem cells might be directly under replenishment by transformed endothelial cells (62).

Our proposition is that the PD-state is initially disseminated in a shorter time-frame through the differentiation process of active stem cell niches. Then, over a distinct slow time-scale on the order of years, the PD-state propagates through stable tissues.

We believe that the hypothesis presented here deserves further investigation and that some experiments should be performed to validate this line of research. To confirm the involvement of circulating progenitor cells in the propagation of PD-state, CD34+ cells (or subpopulations of CD34+ cells) (63) collected from the peripheral blood of PD patients could be transplanted in the bone marrow of nude mice in which all the bone marrow cells are ablated by irradiation prior the transplant. This experiment might give important insights on whether circulating progenitor cells are involved in the etiology of PD. To evaluate whether PD originates at the bone marrow and propagates to the nervous system, or rather initiates at the gastrointestinal tissue, propagates to the nervous system and only via this latter one, reaches the bone marrow, we suggest a long-term, large-scale study where the blood of individuals with ages above 50 years but without PD symptoms would be collected and analyzed over 10 years. If the bone marrow is implicated in the origin of PD, circulating progenitor cells will exhibit a PD-state profile before motor symptoms appear. In contrast, if motor symptoms appear before the PD-state profile is observed in circulating progenitor cells, then the bone marrow is not the origin but rather yet another late stop in the progression of the disease. We chose to highlight one particular hypothetical path from stem-cell ageing to PD, but variant paths cannot be excluded at present. For instance, PD could also originate in the transformation (ageing) of intestinal epithelial stem cells (Lgr5$^+$ cells). Of note, these stem cells have been recently identified as the origin of intestinal cancer (64). In this regard, the additional collection and characterization of intestinal biopsies from the individuals in the aforementioned proposed study would be pertinent to discriminate between these alternate possibilities.

## Materials and Methods

In *Basic algorithm* we describe the overall procedure for building our composite biomarker, based on the testing of RCs. In *Rationality Classes for Parkinson's disease* we describe the RCs that we constructed for this PD problem.

### Basic algorithm

Henceforth, P shall refer to "Parkinson's", H shall refer to "healthy" and A shall refer to "Alzheimer's or other neurological disease". Define a law as a rule that assigns a value to each individual. For example, the expression value of gene GeneX on the individual is a law. The product of the expression values of GeneX, GeneY and GeneZ is another example of a law. The final objective will be to select a law whose values constitute a good diagnostic score for the presence of P. We shall use the convention of equating high values with higher chance of P.

We define the following Law Scoring Function (LSF) to rate how good a law is, based on a set of individuals:

$$\text{LSF score} = \left| \frac{1}{\frac{\sigma_P}{<P> - <A\&H>} + \frac{\sigma_{A\&H}}{<P> - <A\&H>}} \right|,$$

where,

<P> = the average [law value] over the P individuals in the set,
<A&H> = the average over the H and A individuals in the set,
$\sigma_P$ = the standard deviation over the P individuals in the set,
$\sigma_{A\&H}$ = the standard deviation over the H and A individuals in the set.

The LSF score measures, for the set of individuals in question, how well separated is the mean of the P individuals from the mean of the A&H individuals, both in terms of the standard deviation of the P individuals and in terms of the standard deviation of the A&H individuals.

Candidate laws are grouped in RCs, as described in the next section. Based on these RC sets of laws, we now describe in steps how to select a final law:

1. The available set of individuals is randomly divided 40/60 into two sets, set S1 and set S2.
2. The laws in each RC are scored and ranked using the LSF and set S1.
3. A new Super Set of laws is assembled containing the top N ranked laws from each RC [note: due to the small size of the overall dataset in question, it was not possible to further generate a meaningful internal procedure for optimizing N. N=10 was used as a reasonable a priori choice].
4. The laws in the new Super Set are scored and ranked using the LSF and the unused set S2 of patients.
5. A composite law is built. It is built by additive combination of laws to be selected from the top N ranked laws in the Super Set. This selection to belong to the composite law is done as follows: Starting with the top ranked law in the Super Set, and moving down the rank, we test whether adding the next law in question to the composite law increases the LSF score of the current composite law. If it does, then the law is selected and the composite law is updated. In the above process, a law is first normalized via division by its standard deviation in the S2 set. Now, each RC has an RC-counter that counts how many times that RC has contributed with a law to the composite law. Each time a RC contributes with a law, its RC-counter increases by one. The number of members in the final composite law is also recorded. Note that when adding laws as described, care must be taken to add them with the appropriate sign, so that high values always respect the convention of being associated with P.
6. Steps 1 through 5 are repeated, for a new random 40/60 partition into S1 and S2 sets. The RC-counters increase cumulatively throughout these repetitions. The 1 through 5 cycle is repeated until the RCs can be ranked with confidence by their RC-counters (i.e., the ranking approaches the unique ranking in the limit of infinite iterations). The average, over these repeated cycles, of the number of members in the final law is also recorded.

The purpose of the above steps was exclusively to obtain i) the RC-counters ranking of the RCs and ii) the average number of members in the final law. We now turn to actually constructing the final composite law:

7. The RC-counters are all scaled by a common factor. This factor is the largest small enough factor such that after the scaling followed by a subsequent rounding to the nearest integer, the sum of the RC-counters is no greater than the average number of members in the final law, as computed above. The number of laws that RCs contribute to the final composite law are then given by these new, scaled and rounded to the nearest integer, RC-counter values.
8. Using this time the full available set of individuals, the laws in each contributing RC are LSF scored and ranked accordingly. Each RC then contributes to the final composite law the top ranked number of laws determined in step 7 above. Similarly to what occurs in step 5, laws are first normalized via division by their standard deviation, this time in the full set of available individuals. The final composite law is thus built.

To the above procedure there is one significant modification to accommodate what we call one-side biomarkers. This is described next.

<u>One-sided biomarkers</u>
Although the final aim is to build a P vs. non-P classifier, a law that distinguishes well a subset of P individuals from the (rest of P individuals U non-P individuals) would be valuable, as at least it would allow a subset of the P individuals to be identified. Likewise, a law that distinguished well a subset of non-P individuals from the (P individuals U rest of the non-P individuals) set, would be useful too, as it would allow a subset of the non-P individuals to be identified. We call such laws one-side biomarkers, as only on one side of their scale of values do they provide a conclusive diagnostic. We suggest that for complex, heterogeneous pathologies, one-sided biomarkers may turn out to be crucial, beneficially bio-marking and characterizing those pathologies via a diversity of sub-cases. In studies involving a non-extensive number of individuals searching for one-sided biomarkers is difficult, due to insufficient statistical power. However in this P problem we have the advantage of the non-P group being naturally decomposable into two already identified, distinct groups: the H group and the A group.

To the P vs. (A&H) type of biomarker, we add two types of one-sided biomarkers: The H vs. (A&P) one-sided biomarker type and the A vs. (H&P) one-sided biomarker type. Our bookkeeping convention will be to associate a biomarker type to a RC. The algorithm shown in steps above is then modified as follows to accommodate this change:

i) In RCs of type H vs. (A&P) the LSF to be used is

$$\text{LSF score} = \left| \frac{1}{\frac{\sigma_{A\&P}}{<A\&P> - <H>} + \frac{\sigma_H}{<A\&P> - <H>}} \right|,$$

where,

<A&P> = the average over the A and P individuals in the set,
<H> = the average over the H individuals in the set,
$\sigma_{A\&P}$ = the standard deviation over the A and P individuals in the set,
$\sigma_H$ = the standard deviation over the H individuals in the set.

Similarly, in RCs of type A vs. (H&P) the LSF to be used is

$$\text{LSF score} = \left| \frac{1}{\frac{\sigma_{H\&P}}{<H\&P> - <A>} + \frac{\sigma_A}{<H\&P> - <A>}} \right|,$$

where,

<H&P> = the average over the H and P individuals in the set,
<A> = the average over the A individuals in the set,
$\sigma_{H\&P}$ = the standard deviation over the H and P individuals in the set,
$\sigma_A$ = the standard deviation over the A individuals in the set.

ii) In step 3 now three Super Sets are built, each drawing from RCs of one of the above three types.
iii) In step 5, the composite final law is built by testing in turn a law from each of the three internally rank ordered Super Sets. Through this process, the top N laws from each Super Set will be tested. Now, consider the A vs. (H&P) one-sided biomarker. For purposes of our ultimate objective of distinguishing P vs. non-P individuals, this one-sided biomarker is of diagnostic value only on one side of its scale of values, namely on the side identified with A. Suppose the low values are the ones associated with A, while the high values are inconclusive, as they are associated with both H and P. Then, when adding (or testing the addition of) this law to the composite law, rather than adding its value V, we add min (V, midpoint), where midpoint = (<A>+<H&P>)/2 is the midpoint between the averages of the A and (H&P) groups. This way, as it should, only the left side of the scale is of significance in the P vs. non-P composite law. If high values were the ones associated with A, then we should add min(-V, -midpoint), in keeping with our convention of having high values associated with P. Analogous arguments apply to H vs. (A&P) biomarkers and to building the final composite law in step 8.

**Rationality Classes for Parkinson's disease**

We constructed RCs by combining three different factors (Supp. Mat. Figure 1):

1. Group type: First, gene ontology based (13) functional sets of genes were chosen according to existing knowledge or existing predictions that they may play a relevant role in PD. A set of 9 genes that have been genetically linked to PD was also considered. To these, a basic set containing all genes and a set containing only the roughly 1000 most highly expressed genes were added. Finally, protein physical interactions were also considered (65), via two sets of human protein-protein pair-wise interactions (one based on literature curated data (66), the other based on yeast-to-hybrid high-throughput data (66; 67)), and a set of literature curated human protein complexes (11).
2. Biomarker type: This could be the P vs. (A&H) standard biomarker or the A vs. (H&P) or H vs. (A&P) one-sided biomarkers.
3. Mathematical formula: Laws based on single entries, the product of a pair of entries and the log of the ratio of two entries were considered. Further, the single and product cases were both further subdivided into two cases, according to whether the high or the lower value of the law is the one associated with P.

By combining the above three factors, 291 RCs were constructed. For example, the set of laws that try to distinguish A from (P&H), associate single gene high values with (P&H) and where the single gene belongs to the vesicle-mediated transport functional set constitutes one RC.

Remarks:
1. For protein binary interactions only the product and the log ratio of the interacting pair are considered.
2. The entry associated with a protein complex is the product of the expression of its member genes (note this allows products and log ratios amongst complexes to be also considered).
3. For large functional sets, we only evaluated products and log ratios amongst the 200 entries in the set that had the highest LSF scores as single entries. This was for computational time reasons.

4. The highly expressed gene set is not a priori determined and held fixed. This is so in order to maintain proper statistical independence within the algorithm. Hence why we state above that its number of genes is 1000 only approximately.

**Notes**

<u>Computation of the Final Composite Law and the associated area under the curve (AUC) with confidence interval</u>

When computing the final composite law, 5 000 iterations of the step 1 through step 5 cycle in the algorithm description above were performed. The top 50 ranked by RC-counter RCs are shown in Supp. Mat. Table 1. The number of patients in the sets S1 and S2 also alluded to in the algorithm description above was 42 and 63, respectively.

To determine the associated area under the curve (AUC), an extra external cycle, leaving 5 random patients out at time, was created. The number of patients in the sets S1 and S2 became 40 and 60, respectively. The outer testing cycle was repeated 500 times and the iterations of the inner step 1 through step 5 cycle were reduced to 50, for computational time reasons.

When determining the standard deviation confidence interval for the AUC estimate, two approximations were made. First, liberally, we assume we have approximately tested 105 independent patients. Second, conservatively, we consider only the minimum number of guaranteed independent (PD patient, non-PD patient) pairs amongst these 105 patients (68; 69). This minimum number is min(50,55)=50, since there are 55 PD patients and 50 non-PD patients. The standard deviation in the AUC estimate thus becomes $sqrt[AUC*(1-AUC)/50] \leq sqrt[0.25/50] \approx 0.07$.

<u>Datasets</u>

The blood genome-wide expression dataset on Parkinson's patients and controls was obtained from Scherzer et al. (3). The data was used as provided in the NCBI Geo Gene Expression Omnibus (70), without any further normalization. The literature curated human protein-protein pair-wise interactions dataset was obtained from Rual et al. (66). The Y2H based high-throughput human protein-protein pair-wise interactions dataset consists of the union of the Rual et al. (66) and the Stelzl et al. (67) datasets. The literature curated human protein complexes dataset was obtained from Ruepp et al. (11).

<u>Go annotations</u>

The functional groups used when constructing Rationality Classes were based on the Gene Ontology classification scheme (13), as given in the Affymetrix Human Genome U133 Array data sheet of November 2009.

**Aknowledgements**

TFO is supported by an EMBO installation grant, a Marie Curie International Reintegration Grant and FTC Grant PIC/IC/82760/2007.

**a)**

| | Functional group | One-sided biomarker type | Expression sign | Genes and mathematical formula | Affymetrix HG-U133A probes |
|---|---|---|---|---|---|
| 1 | Vesicle-mediated transport | **P vs. (A & H)** | High in P | CLTB | 211043_s_at |
| 2 | G-protein coupled receptors | **P vs. (A & H)** | High in P | FPR3 | 214560_at |
| 3 | Human literature protein complexes | **A vs (P & H)** | High in A | CD47 * ITGA2 * ITGB1 | (211075_s_at * 213055_at * 213856_at * 213857_s_at) * (205032_at) * (211945_s_at * 215878_at * 215879_at * 216178_x_at * 216190_x_at) |
| 4 | Ubiquitination | **A vs (P & H)** | Low in A | UBOX5 * PRPF19 | 215544_s_at * 203103_s_at |
| 5 | Vesicle-mediated transport | **P vs. (A & H)** | High in P | CLTB * KIF5B | 211043_s_at * 201991_s_at |
| 6 | Inflammatory | **H vs. (A & P)** | Low in H | PTGS2 * CYBB | 204748_at * 217431_x_at |
| 7 | Oxidative stress | **A vs (P & H)** | Low in A | GPX4 | 201106_at |
| 8 | Dopamine | **A vs (P & H)** | High in A | PTGS2 | 204748_at |
| 9 | Genetically linked to Parkinson | **H vs. (A & P)** | Low in H | UCHL1 | 201387_s_at |
| 10 | Folding | **H vs. (A & P)** | High in H | ST13 | 208666_s_at |
| 11 | Nitric Oxide | **A vs (P & H)** | High in A | ACTB | AFFX-HSAC07/X00351_M_at |
| 12 | Literature physically interacting protein pairs | **H vs. (A & P)** | High in H | SMAD3 * STRAP | 205398_s_at * 200870_at |
| 13 | Ubiquitination | **P vs. (A & H)** | High in P | UBOX5 * CUL4B | 215544_s_at * 202213_s_at |
| 14 | Ubiquitination | **H vs. (A & P)** | High in H | SMAD3 | 205398_s_at |

**b)**

| Parkinson's disease composite final law |
|---|
| + CLTB / 55.7 |
| + FPR3 / 28 |
| + min (- CD47*ITGA2*ITGB1, - 1.806*10$^{18}$) / (3.131*10$^{18}$) |
| + min (UBOX5*PRPF19, 8023.4) / 5916.2 |
| + CLTB*KIF5B / 16654 |
| + min (PTGS2*CYBB, 1109.5) / 1327.6 |
| + min (GPX4, 544.1) / 124.9 |
| + min (- PTGS2, - 68.1) / 36.2 |
| + min (UCHL1, 81.2) / 42.5 |
| + min (- ST13, - 90.5) / 25.4 |
| + min (- ACTB, - 5761.7) / 1420.9 |
| + min ( - SMAD3*STRAP, - 16611.8) / 8477.5 |
| + UBOX5*CUL4B / 4221.5 |
| + min (- SMAD3, - 50.5) / 24.9 |

Table 1. a) The 14 laws that make up the final composite Parkinson's disease biomarker and their respective Rationality Class of origin. b) The composite gene-expression biomarker for Parkinson's disease. Gene symbols stand for their expression level. For normalization, every term is divided by its standard deviation.

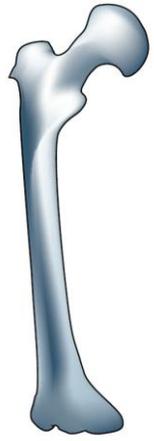

Hematopoietic stem cells

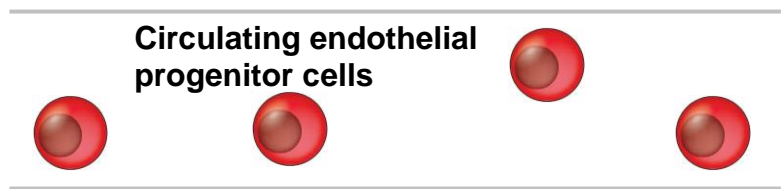

Circulating endothelial progenitor cells

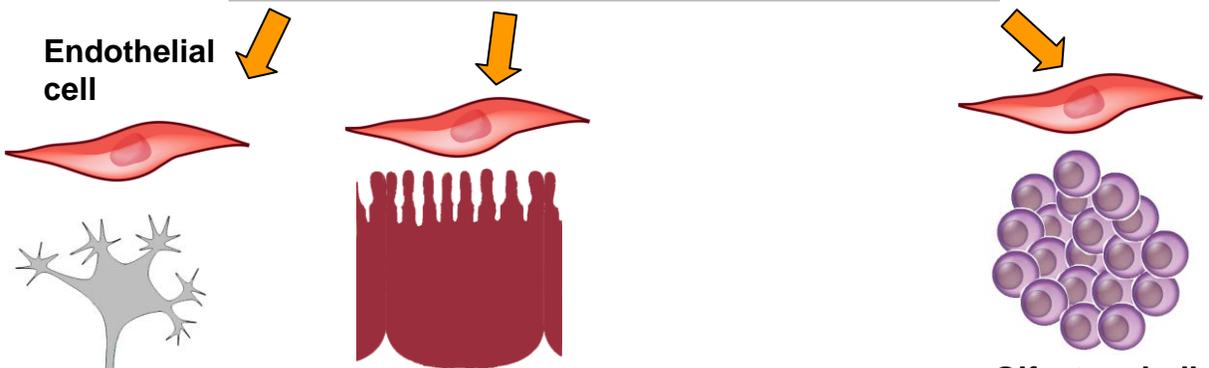

Endothelial cell

Epithelial stem cells

Olfactory bulb

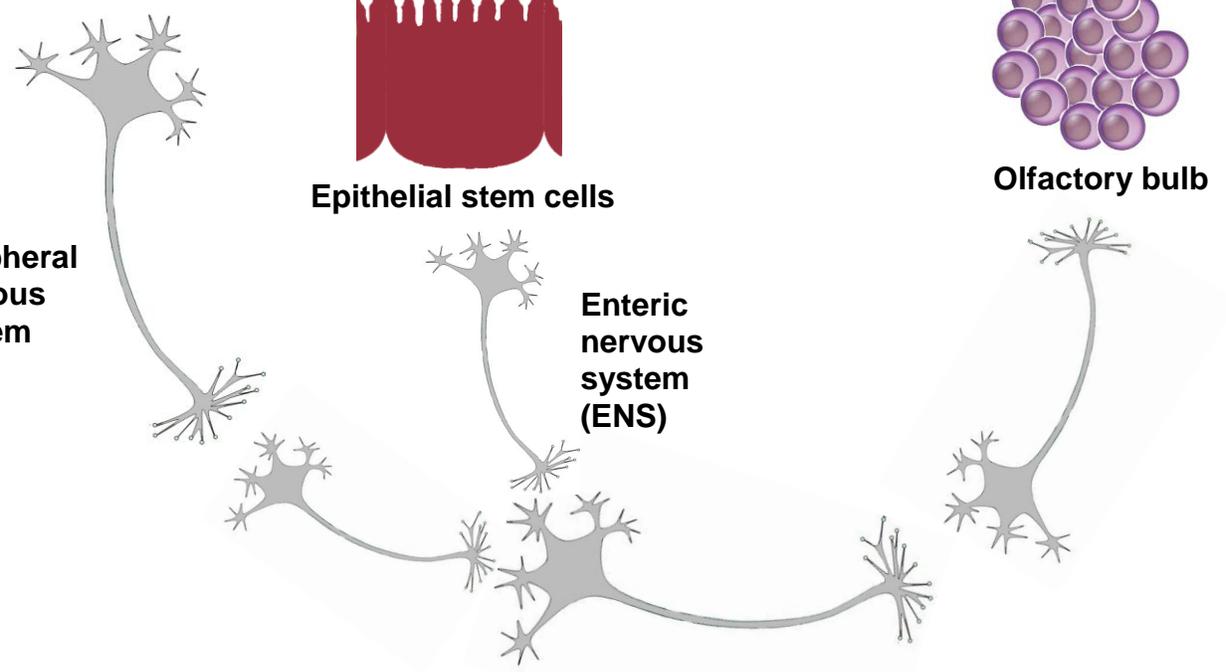

Peripheral nervous system

Enteric nervous system (ENS)

Central nervous system

Figure 1. Bone marrow-derived stem cells as the origin of PD. According to this hypothesis, the PD-state originally appears in bone marrow-derived stem cells due to ageing. Bone marrow-resident hematopoietic stem cells give rise to blood cells in a PD-state, including circulating endothelial progenitor cells (CEPCs). CEPCs differentiate into endothelial cells which are incorporated into blood vessels, hence propagating the disease at different sites of the human body. The first symptoms of the disease (as early as 10 years before motor symptoms) would occur at places where the cell turnover is high, for instance at gastrointestinal tissue and at the olfactory bulb.

| | One-sided biomarker type |
| --- | --- |
| | (P=Parkinson, H=healthy, A=Alzheimer or other neurological disorder) |
| | P vs. (A & H) |
| | H vs. (A & P) |
| | A vs. (H & P) |

| Group type | Group Name | # of probes |
| --- | --- | --- |
| Functional | Apoptosis and cell death | 1896 |
| | Dopamine | 119 |
| | Folding | 255 |
| | G-protein coupled receptors | 809 |
| | Immune system | 1014 |
| | Inflammation | 547 |
| | Iron or heme homeostasis | 565 |
| | Lipid metabolism/catabolism | 595 |
| | Matrix metalloproteinases | 45 |
| | Metabolism | 2823 |
| | Mitochondria | 1516 |
| | Nitric Oxide | 145 |
| | Oxidative stress | 199 |
| | Ubiquitination | 643 |
| | Vesicle-mediated transport | 402 |
| Specific | Genetically linked to Parkinson | 9 |
| Generic | All genes | 22283 |
| | High-expression genes | ☐1000 |
| Protein interactions | Y2H based physically interacting protein pairs | 10169 pairs |
| | Literature physically interacting protein pairs | 14082 pairs |
| | Literature human protein complexes | 1266 |

| | Total # of Rationality Classes: 291 |
| --- | --- |

| Mathematical formula | Expression sign (P=Parkinson, H=healthy, A=Alzheimer or other neurological disorder) |
| --- | --- |
| Single entry | Higher expression in P (or in H; or in A, depending on one-sided biomarker type) |
| Single entry | Lower expression in P (or in H; or in A, depending on one-sided biomarker type) |
| Product of 2 entries | Higher expression product in P (or in H; or in A, depending on one-sided biomarker type) |
| Product of 2 entries | Lower expression product in P (or in H; or in A, depending on one-sided biomarker type) |
| Log (ratio of 2 entries) | High absolute log(ratio) score |

**Supp. Mat. Figure 1.** The construction of Rationality Classes for Parkinson's disease genome-wide expression analysis.

| | Class Name | Class Type | Expression Sign | Operation Type | RC-counter (after 5000 iterations) |
|---|---|---|---|---|---|
| 1 | Vesicle-mediated transport | P vs. (A & H) | High in P | Simple | 2880 |
| 2 | G-protein coupled receptors | P vs. (A & H) | High in P | Simple | 1989 |
| 3 | Literature human protein complexes | A vs (P & H) | High in A | Simple | 1642 |
| 4 | Ubiquitination | A vs (P & H) | Low in A | Product | 1624 |
| 5 | Vesicle-mediated transport | P vs. (A & H) | High in P | Product | 1390 |
| 6 | Inflammation | H vs. (A & P) | Low in H | Product | 1345 |
| 7 | Oxidative stress | A vs (P & H) | Low in A | Simple | 1296 |
| 8 | Dopamine | A vs (P & H) | High in A | Simple | 1232 |
| 9 | Genetically linked to Parkinson | H vs. (A & P) | Low in H | Simple | 1182 |
| 10 | Folding | H vs. (A & P) | High in H | Simple | 1137 |
| 11 | Nitric Oxide | A vs (P & H) | High in A | Simple | 1126 |
| 12 | Literature physically interacting protein pairs | H vs. (A & P) | High in H | Product | 1101 |
| 13 | Ubiquitination | P vs. (A & H) | High in P | Product | 976 |
| 14 | Ubiquitination | H vs. (A & P) | High in H | Simple | 975 |
| 15 | Nitric Oxide | A vs (P & H) | Low in A | Simple | 972 |
| 16 | Immune system | H vs. (A & P) | High in H | Simple | 957 |
| 17 | Literature human protein complexes | H vs. (A & P) | High in H | Simple | 932 |
| 18 | Oxidative stress | A vs (P & H) | High in A | Simple | 905 |
| 19 | Dopamine | H vs. (A & P) | Low in H | Simple | 885 |
| 20 | Iron or heme homeostasis | H vs. (A & P) | Low in H | Product | 872 |
| 21 | Ubiquitination | H vs. (A & P) | High in H | Product | 852 |
| 22 | Dopamine | A vs (P & H) | High in A | Product | 846 |
| 23 | High-expression genes | A vs (P & H) | Low in A | Simple | 758 |
| 24 | High-expression genes | P vs. (A & H) | High in P | Simple | 724 |
| 25 | Literature physically interacting protein pairs | H vs. (A & P) | High absolute | Ratio Log | 721 |
| 26 | Folding | H vs. (A & P) | High absolute | Ratio Log | 665 |
| 27 | Genetically linked to Parkinson | P vs. (A & H) | High in P | Simple | 659 |
| 28 | Iron or heme homeostasis | A vs (P & H) | High in A | Simple | 651 |
| 29 | High-expression genes | A vs (P & H) | Low in A | Product | 599 |
| 30 | Iron or heme homeostasis | A vs (P & H) | High absolute | Ratio Log | 596 |
| 31 | Inflammation | A vs (P & H) | Low in A | Simple | 566 |
| 32 | All genes | P vs. (A & H) | High in P | Simple | 560 |
| 33 | Mitochondria | H vs. (A & P) | High in H | Product | 555 |
| 34 | Oxidative stress | H vs. (A & P) | Low in H | Simple | 531 |
| 35 | Literature human protein complexes | P vs. (A & H) | Low in P | Simple | 507 |
| 36 | Iron or heme homeostasis | A vs (P & H) | High in A | Product | 500 |
| 37 | G-protein coupled receptors | P vs. (A & H) | High in P | Product | 499 |
| 38 | Ubiquitination | A vs (P & H) | Low in A | Simple | 483 |
| 39 | Apoptosis and cell death | H vs. (A & P) | High absolute | Ratio Log | 436 |
| 40 | Mitochondria | P vs. (A & H) | High absolute | Ratio Log | 431 |
| 41 | High-expression genes | P vs. (A & H) | High in P | Product | 428 |
| 42 | Literature human protein complexes | H vs. (A & P) | Low in H | Simple | 407 |
| 43 | Matrix metalloproteinases | P vs. (A & H) | Low in P | Product | 397 |
| 44 | Nitric Oxide | A vs (P & H) | Low in A | Product | 368 |
| 45 | Inflammation | A vs (P & H) | Low in A | Product | 364 |
| 46 | Mitochondria | P vs. (A & H) | High in P | Simple | 361 |
| 47 | Folding | P vs. (A & H) | High absolute | Ratio Log | 359 |
| 48 | Apoptosis and cell death | A vs (P & H) | Low in A | Product | 354 |
| 49 | Mitochondria | H vs. (A & P) | High in H | Simple | 347 |
| 50 | Ubiquitination | P vs. (A & H) | High in P | Simple | 342 |

**Supp. Mat. Table 1.** The top 50 Rationality Classes, rank ordered by RC-counter.